\documentclass[conference]{IEEEtran}
\IEEEoverridecommandlockouts

\usepackage{cite}
\usepackage{amsmath,amssymb,amsfonts}
\usepackage{algorithmic}
\usepackage{graphicx}
\usepackage{textcomp}
\usepackage{xcolor}
\usepackage{booktabs}
\usepackage{multirow}
\usepackage{comment}
\def\BibTeX{{\rm B\kern-.05em{\sc i\kern-.025em b}\kern-.08em
    T\kern-.1667em\lower.7ex\hbox{E}\kern-.125emX}}
\begin{document}

\title{Who Said What (WSW~2.0)? Enhanced Automated Analysis of Preschool Classroom Speech\\
}

\author{\IEEEauthorblockN{Anchen Sun\IEEEauthorrefmark{1},
Tiantian Feng\IEEEauthorrefmark{2},
Gabriela Gutierrez\IEEEauthorrefmark{3},
Juan J Londono\IEEEauthorrefmark{3}, 
Anfeng Xu\IEEEauthorrefmark{2}, \\
Batya Elbaum\IEEEauthorrefmark{4},
Shrikanth Narayanan\IEEEauthorrefmark{2},
Lynn K Perry\IEEEauthorrefmark{3},
Daniel S Messinger\IEEEauthorrefmark{3}
\IEEEauthorblockA{\IEEEauthorrefmark{1}Department of Electrical and Computer Engineering\\University of Miami, Coral Gables,
FL, USA}
\IEEEauthorblockA{\IEEEauthorrefmark{2}Department of Computer Science, Viterbi School of Engineering\\University of Southern California, Los Angeles, CA, USA}
\IEEEauthorblockA{\IEEEauthorrefmark{3}Department of Psychology\\University of Miami, Coral Gables,
FL, USA}
\IEEEauthorblockA{\IEEEauthorrefmark{4}Department of Teaching and Learning \\University of Miami, Coral Gables,
FL, USA \\
Email: asun@miami.edu, tiantiaf@usc.edu, gag160@miami.edu, jjl282@miami.edu, anfengxu@usc.edu, \\ elbaum@miami.edu, shri@usc.edu, lkperry@miami.edu, dmessinger@miami.edu}
}} 

\maketitle

\begin{abstract}
 This paper introduces an automated framework (WSW~2.0) for analyzing vocal interactions in preschool classrooms, enhancing both accuracy and scalability through the integration of wav2vec2-based speaker classification and Whisper (large-v2 and large-v3) speech transcription. A total of 235 minutes of audio recordings (160 minutes from 12 children and 75 minutes from 5 teachers), were used to compare system outputs to expert human annotations. WSW~2.0 achieves a weighted F1 score of .845, accuracy of .846, and an error-corrected kappa of .672 for speaker classification (child vs. teacher). Transcription quality is moderate to high with word error rates of .119 for teachers and .238 for children. WSW~2.0 exhibits relatively high absolute agreement intraclass correlations (ICC) with expert transcriptions for a range of classroom language features. These include teacher and child mean utterance length, lexical diversity, question asking, and responses to questions and other utterances, which show absolute agreement intraclass correlations between .64 and .98. To establish scalability, we apply the framework to an extensive dataset spanning two years and over 1,592 hours of classroom audio recordings, demonstrating the framework’s robustness for broad real-world applications. These findings highlight the potential of deep learning and natural language processing techniques to revolutionize educational research by providing accurate measures of key features of preschool classroom speech, ultimately guiding more effective intervention strategies and supporting early childhood language development. 
\end{abstract}


\section{Background}

\subsection{Overview} Preschool classrooms are critical environments for developing core communication skills in young children. Our previous work, WSW~1.0~\cite{WSW1}, introduced an automated framework designed to analyze linguistic interactions in preschool classrooms and demonstrated accurately transcribed classroom conversations with a relatively low word error rate. Building on this foundation, the current study presents \textbf{WSW~2.0}, a framework for large-scale analysis of classroom interactions, incorporating a wav2vec2-based speaker classification model and Whisper transcription models. WSW 2.0 reliability was assessed with more than two times the quantity of annotated audio from teacher and child recordings used to assess WSW 1.0. These data, combined with advanced deep learning architectures for speech recognition and speaker classification, allow WSW~2.0 to address real-world challenges in preschool language assessment.

\subsection{Advances in Speaker Classification and Speech Recognition} Identifying whether an utterance is produced by a child or teacher is important for accurately transcribing teacher and child speech. An egocentric speaker classification (ESC(wav2vec2-based)) model~\cite{feng2024egocentric}, provides robust performance and scalability for this task. Commercial products such as LENA (Language Environment Analysis)~\cite{gilkerson2017mapping} are widespread, and previous efforts, such as VTC (Voice Type Classifier) for child diarization, have also shown promising results~\cite{lavechin20_interspeech}. However, our pipeline broadens accessibility by optimizing open-source solutions. Meanwhile, significant strides in speech recognition, fueled by end-to-end systems such as OpenAI's Whisper, have dramatically improved transcription capabilities under varying speech and speaker conditions~\cite{graham2024evaluating}. These advances enable more accurate large-scale analyses of teacher–child interactions than were available even a few years ago. However, performance gaps between automated processing of adult and child speech remain, suggesting the need for new child-specific algorithmic enhancements to further improve performance~\cite{shivakumar2022end}.

\subsection{Whisper} Whisper is a deep learning-based automatic speech recognition (ASR) model trained on massive audio–text corpora
. Recent releases, including the large-v2 and large-v3 variants, have demonstrated significant gains in transcription accuracy for adult speech~\cite{radford2023robust}. However, preschool environments present additional complexity due to diverse speaker profiles and background noise. We systematically evaluate Whisper large-v2 and large-v3 [both in combination with ESC(wav2vec2)] to improve speaker classification and transcription accuracy. 

\section{Current Paper}

WSW~2.0 leverages modern speaker classification (ESC(wav2vec2)) and  speech recognition software (Whisper large-v2) to  capture teacher–child interactions. We have  expanded our reliability dataset—introducing more teachers, more children, and nearly three times as much annotated audio—as used in WSW 1.0. Within this larger framework, we quantify a range of features linked to developmental outcomes: teacher mean length of utterance (MLU), teacher lexical diversity, frequency of teacher and child questions, teacher–child responsivity, and child MLU. Below we motivate our choice of structural (MLU and lexical diversity) and responsivity-oriented (question asking and responsivity) language indices in the current paper.

 The mean length of a speaker’s utterances in words (MLU) is associated with lexical diversity, i.e., the number of different words in an utterance or recording, as well as morphosyntactic complexity~\cite{hadley2022meta, potratz2022measurement}
 , making it a robust index of expressive language development for both typically developing children and children with disabilities~\cite{flipsen2014mean}
 . In preschool environments, higher teacher MLU and lexical diversity are both associated with gains in measured language ability in both typically developing children and those with disabilities~\cite{huttenlocher2002language, justice2013bi, yang2022measurement}
 . Indeed, the MLU of the adult language children hear shows stronger associations with their language outcomes than the quantity of adult utterances that they hear~\cite{anderson2021linking}.

Teacher questions can play a key role in children’s language development, primarily by eliciting  longer and more complex child utterances~\cite{hadley2022meta, zucker2020asking}. 
Children’s question-asking has also been associated with later language and academic outcomes~\cite{kurkul2022you, ronfard2018question}. 
Higher caregiver responsiveness to children’s utterances is associated with greater syntactic and lexical abilities in both typically developing children~\cite{donnelly2021longitudinal, elmlinger2019ecology} 
and children with disabilities~\cite{bottema2018sequential, clark2022caregiver} 
, and both teacher and child responsivity contribute to children’s language development~\cite{cabell2015teacher, justice2018linguistic}.

\section{Proposed Framework}

\begin{figure}
    \centering
    \includegraphics[width=0.48\textwidth]{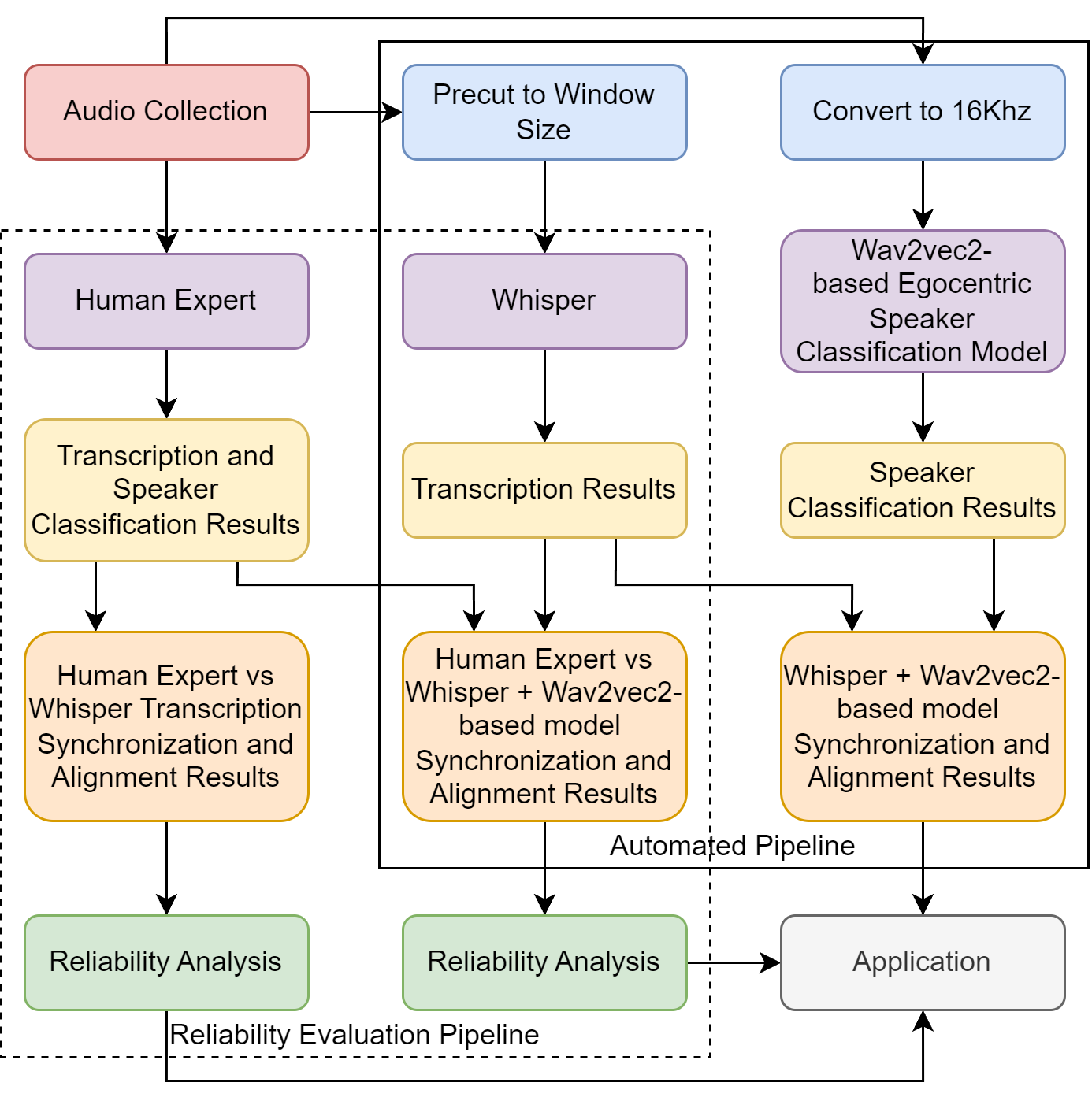}
    \caption{Workflow. Whisper is automated speech-to-text (transcription) software. The ESC(wav2vec2) provides automated speaker classification (teacher versus child). The human expert provides both speaker classification and transcription. Whisper transcription is used to synchronize the ESC(wav2vec2) and expert speaker classification.}
    \label{fig:dia}
\end{figure}

Figure~\ref{fig:dia} outlines the WSW~2.0 framework for large-scale teacher–child vocal interaction analysis. The ESC(wav2vec2) speaker classifier is integrated with Whisper (either large-v2 or large-v3), and is systematically aligned with expert annotations for validation. We additionally report reliability analyses demonstrating WSW~2.0’s relative accuracy in quantifying preschool language.

\section{Application}

The WSW 2.0 framework was implemented using Python 3.10.16 with PyTorch 2.5.1, CUDA 12.4, and cuDNN 9.1 on an Ubuntu 22.04 operating system. 

\subsection{Data Collection and Processing}
Audio recordings were collected twice monthly in a preschool classroom across the school year, with each session lasting about three hours and twenty minutes on average. Each child and teacher wore a Sony ICD-UX570 stereo recorder (LPCM 44.1 kHz/16-bit) in a vest (for children) or a fanny pack (for teachers). Informed consent was obtained from all teachers and parents or guardians. 


This study evaluated reliability through two tasks: speaker classification (teacher versus child) and language feature extraction, which involves transcription. A total of 235 minutes of audio recordings (160 minutes from 12 children and 75 minutes from 5 teachers), were used to compare system outputs to expert human annotations. The speaker classification task used 145 minutes of audio (85 from child recorders, 60 from teacher recorders), involving 9 children (5 girls) aged 2.8–4.6 years, as well as 4 teachers, from 6 classrooms across 2 academic years. One teacher contributed 3 recordings across 2 classes. The 90 minutes of audio recordings not used for the speaker classification task (235 - 145) included 20 minutes of audio for which analyses have not yet been performed and 70 minutes for which experts had not classified speakers (data were not available).  
The language feature task (including transcription) used 210 minutes of audio (155 from children, 55 from teachers), involving 11 children (6 girls) aged 2.8–4.6 years, as well as 3 teachers, from 4 classrooms over 2 academic years. The 210 minutes of language feature recordings did not include 25 minutes of classrooms for which analyses have not been performed. The 210 minutes of language feature recordings included 75 minutes coded by experts by listening to the audio file and editing a copy of the Whisper transcription to indicate disagreements~\cite{liu2023automation}. The additional 135 minutes were coded using only the audio files.


\subsection{Speaker Classification Reliability}

\begin{table}[h]
    \centering
    \begin{tabular}{|cc|c|c|c|}
    \hline
    \multicolumn{1}{|c|}{Year\&Class} & Recorder  & F1 Score & Accuracy & Cohen's Kappa \\ \hline
    \multicolumn{1}{|c|}{2223 C1}     & Child 1   & 0.907    & 0.907    & 0.799         \\ \hline
    \multicolumn{1}{|c|}{2223 C1}     & Child 2   & 0.741    & 0.756    & 0.316         \\ \hline
    \multicolumn{1}{|c|}{2223 C2}     & Child 3   & 0.784    & 0.782    & 0.568         \\ \hline
    \multicolumn{1}{|c|}{2223 C3}     & Child 4   & 0.909    & 0.907    & 0.740         \\ \hline
    \multicolumn{1}{|c|}{2223 C3}     & Child 5   & 0.826    & 0.820    & 0.620         \\ \hline
    \multicolumn{1}{|c|}{2223 C3}     & Child 6   & 0.839    & 0.840    & 0.668         \\ \hline
    \multicolumn{1}{|c|}{2223 C3}     & Child 7   & 0.828    & 0.826    & 0.642         \\ \hline
    \multicolumn{1}{|c|}{2223 C3}     & Teacher 1 & 0.885    & 0.884    & 0.697         \\ \hline
    \multicolumn{1}{|c|}{2223 C2}     & Teacher 2 & 0.863    & 0.862    & 0.687         \\ \hline
    \multicolumn{1}{|c|}{2223 C3}     & Teacher 1 & 0.821    & 0.825    & 0.628         \\ \hline
    \multicolumn{1}{|c|}{2223 C4}     & Teacher 3 & 0.880    & 0.885    & 0.574         \\ \hline
    \multicolumn{1}{|c|}{2324 C5}     & Child 8   & 0.877    & 0.865    & 0.568         \\ \hline
    \multicolumn{1}{|c|}{2324 C5}     & Child 9   & 0.848    & 0.850    & 0.699         \\ \hline
    \multicolumn{1}{|c|}{2324 C5}     & Teacher 1 & 0.882    & 0.885    & 0.760         \\ \hline
    \multicolumn{1}{|c|}{2324 C6}     & Teacher 4 & 0.818    & 0.817    & 0.593         \\ \hline
    \multicolumn{2}{|c|}{Time-Weighted Mean}           & 0.845    & 0.847    & 0.635         \\ \hline
    \multicolumn{2}{|c|}{Overall}                 & 0.845    & 0.846    & 0.672         \\ \hline
    \end{tabular}
    \caption{Human coding vs WSW 2.0 (Whisper large-v2 + ESC(wav2vec2)).}
    \label{tab:performance_metrics}
\end{table}

\begin{figure}
    \centering
    \includegraphics[width=0.3\textwidth]{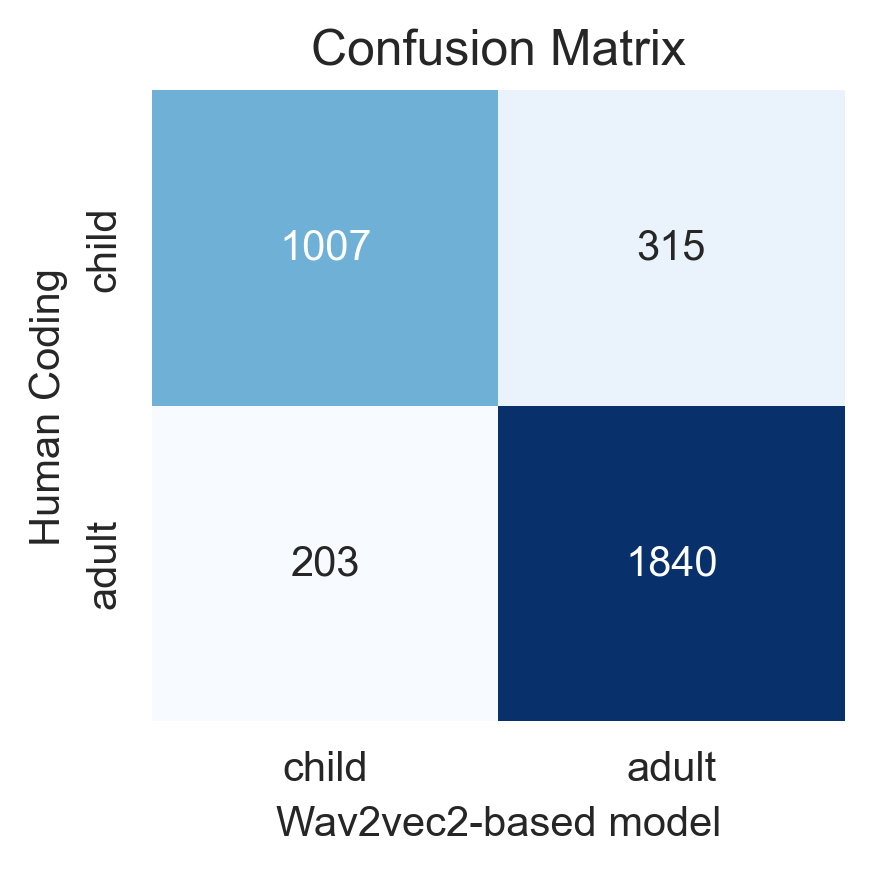}
    \caption{Cross-classification confusion matrix between Whisper large-v2 + ESC(wav2vec2) and the expert. The ESC(wav2vec2) produced the classification (teacher versus child) while Whisper transcription was used to align results with the expert transcription. Teacher utterances were classified more accurately than child utterances (see TABLE~\ref{tab:performance_metrics}, Overall, for statistical descriptions)}
    \label{fig:conf}
\end{figure}

TABLE~\ref{tab:performance_metrics} shows utterance-level reliability metrics for our integrated system Whisper large-v2 plus ESC(wav2vec2) compared to expert annotations, where ESC stands for egocentric speaker classification. Analyzing 210 minutes of data, we achieved an overall accuracy of 0.846, with a weighted F1 score of 0.845 and a Cohen’s kappa of 0.672 (adjusted for chance, 0.635 for time-weighted mean). Fig.~\ref{fig:conf} displays the confusion matrix, which indicates notably higher precision for teacher utterances than for child utterances, although both speaker groups are classified well.

\subsection{Transcription and language feature Reliability}

\begin{table}[]
\begin{tabular}{|l|ll|ll|ll|}
\hline
\multirow{2}{*}{ICCs} & \multicolumn{2}{l|}{Overall MLU}                    & \multicolumn{2}{l|}{Question MLU}                  & \multicolumn{2}{l|}{Non-Question MLU}                \\ \cline{2-7} 
                      & \multicolumn{1}{l|}{Teacher}        & Child         & \multicolumn{1}{l|}{Teacher}        & Child        & \multicolumn{1}{l|}{Teacher}        & Child          \\ \hline
large-v2              & \multicolumn{1}{l|}{\textbf{0.986}} & \textbf{0.770} & \multicolumn{1}{l|}{0.940}           & \textbf{0.800} & \multicolumn{1}{l|}{\textbf{0.978}} & \textbf{0.813} \\ \hline
large-v3              & \multicolumn{1}{l|}{0.985}          & 0.653         & \multicolumn{1}{l|}{\textbf{0.963}} & 0.281        & \multicolumn{1}{l|}{0.951}          & 0.659          \\ \hline
\end{tabular}
\caption{Intraclass Correlations Comparison between integrated Whisper large-v2 and large-v3 in WSW (Whisper vs. Expert).}
\label{table:whisper}
\end{table}

We compared Whisper large-v2 and large-v3 for their impact on speech-feature extraction 
by comparing their outputs with expert annotations at a file-
level granularity. To do so, we computed absolute
agreement intraclass correlations (ICCs) using audio recordings as the unit of analysis. These absolute
agreement ICCs indicate the proportion of total variance
attributable to true differences across audio files rather than
variance caused by measurement error or systematic discrep-
ancies between systems. Large-v3 yielded lower ICCs than large-v2. This was particularly evident for child mean length of utterance (MLU) including the mean length of child question utterances (0.281 from large-v3 vs. 0.800 from large-v2) as shown in TABLE~\ref{table:whisper}. Consequently, we chose large-v2 for the final WSW~2.0 pipeline.

\begin{table}[]
\centering
\begin{tabular}{|l|l|l|l|l|l|l|}
\hline
   & \begin{tabular}[c]{@{}l@{}}Whisper\\ Transcription\end{tabular}                       & \begin{tabular}[c]{@{}l@{}}Expert \\ Transcription\end{tabular}                      & T/C & \begin{tabular}[c]{@{}l@{}}Expert \\ Word \\ Count\end{tabular} & LD & WER  \\ \hline
1  & \begin{tabular}[c]{@{}l@{}}How is the \\ weather?\end{tabular}                        & \begin{tabular}[c]{@{}l@{}}How is the \\ weather?\end{tabular}                       & T   & 4                                                               & 0  & 0    \\ \hline
2  & Sunny.                                                                                & Sunny                                                                                & C   & 1                                                               & 0  & 0    \\ \hline
3  & Sunny.                                                                                & It's rainy?                                                                          & T   & 2                                                               & 2  & 1    \\ \hline
4  & \begin{tabular}[c]{@{}l@{}}It's sunny? \\ I don't know \\ if it's sunny.\end{tabular} & \begin{tabular}[c]{@{}l@{}}It's sunny? \\ I don't know \\ if it's sunny\end{tabular} & T   & 8                                                               & 0  & 0    \\ \hline
5  & It is.                                                                                & It is                                                                                & C   & 2                                                               & 0  & 0    \\ \hline
6  & \begin{tabular}[c]{@{}l@{}}It is? I think \\ it's sunny \\ as well.\end{tabular}      & \begin{tabular}[c]{@{}l@{}}It is? I think \\ it's sunny \\ as well.\end{tabular}     & T   & 8                                                               & 0  & 0    \\ \hline
7  & Me too.                                                                               & Me too.                                                                              & C   & 2                                                               & 0  & 0    \\ \hline
8  & \begin{tabular}[c]{@{}l@{}}Yeah, you \\ too.\end{tabular}                             & \begin{tabular}[c]{@{}l@{}}Yeah, you \\ too.\end{tabular}                            & T   & 3                                                               & 0  & 0    \\ \hline
9  & Oh, raisins.                                                                          & \begin{tabular}[c]{@{}l@{}}Oh a raisin \\ is in there\end{tabular}                   & C   & 6                                                               & 5  & 0.83 \\ \hline
10 & \begin{tabular}[c]{@{}l@{}}I love the \\ raisins.\end{tabular}                        & \begin{tabular}[c]{@{}l@{}}I love the \\ raisins\end{tabular}                        & C   & 4                                                               & 0  & 0    \\ \hline
\end{tabular}
\caption{Transcription Comparison. T/C-Teacher or Child; LD-Levenshtein Distance; WER-word error rate(LD/Word Count).}
\label{table:wer}
\end{table}

\subsubsection{Word Error Rate (WER)} As a stringent benchmark, we computed WER separately for teacher and child speech, using segments where the assigned speaker matched the microphone (child-worn or teacher-worn). WER is determined using the Levenshtein Distance, which measures the difference between two utterances by counting the number of word modifications required to transform one into the other (see TABLE~\ref{table:wer}). Any utterance not detected by Whisper or the expert is treated as having all words incorrect. With Whisper large-v2 in WSW~2.0, the mean WER is 0.119 for teacher speech and 0.238 for child speech, indicating that 88\% of teacher words and 76\% of child words align with expert transcripts. These values suggest the pipeline’s potential for large-scale classroom audio analysis.

\subsubsection{Language Feature Comparison Overview} Beyond word-level accuracy, WSW 2.0 also quantifies a range of language features critical to classroom discourse analysis. We examined teacher and child speech extracted from both teacher- and child-worn microphones to capture 
Whisper- and expert-derived estimates of words per minute, MLU, number of utterances, frequency of questions and non-questions, the proportion of utterances that were responded to, and the mean lexical diversity (rate of unique words per minute). Utterances formed the primary unit of analysis for computing descriptive statistics and correlations (see TABLE~\ref{table:large}).



\begin{table*}[h]
\centering
\setlength\tabcolsep{5pt}
\begin{tabular}{|cccccccccccc|}
\hline
\multicolumn{1}{|c|}{}        & \multicolumn{2}{c|}{Child Words}                                                                                        & \multicolumn{3}{c|}{Child Utterances}                                                                                                                                                                        & \multicolumn{2}{c|}{Teacher Responses}                                                                                                                          & \multicolumn{2}{c|}{\begin{tabular}[c]{@{}c@{}}Proportion of \\ Child Utterances \\ Followed by a \\ Teacher Response\end{tabular}} & \multicolumn{1}{c|}{\begin{tabular}[c]{@{}c@{}} Percentage\end{tabular}} & \multicolumn{1}{c|}{\begin{tabular}[c]{@{}c@{}} Lexical \\Diversity  \end{tabular}}                                                                                                                      \\ \hline
\multicolumn{1}{|c|}{}        & \multicolumn{1}{c|}{MLU}  & \multicolumn{1}{c|}{\begin{tabular}[c]{@{}c@{}}Mean \\ Words \\ per Minute\end{tabular}} & \multicolumn{1}{c|}{\begin{tabular}[c]{@{}c@{}}Total \\ Child \\ Utterances\end{tabular}}   & \multicolumn{1}{c|}{Questions} & \multicolumn{1}{c|}{\begin{tabular}[c]{@{}c@{}}Non-\\ Questions\end{tabular}} & \multicolumn{1}{c|}{\begin{tabular}[c]{@{}c@{}}To \\ Questions\end{tabular}} & \multicolumn{1}{c|}{\begin{tabular}[c]{@{}c@{}}To Non-\\ Questions\end{tabular}} & \multicolumn{1}{c|}{Questions}            & \multicolumn{1}{c|}{\begin{tabular}[c]{@{}c@{}}Non-\\ Questions\end{tabular}}           & \multicolumn{1}{c|}{Questions}                                                                                            & \begin{tabular}[c]{@{}c@{}}Mean Lexical\\Diversity\\per Minute\end{tabular}  \\ \hline
\multicolumn{1}{|c|}{Whisper} & \multicolumn{1}{c|}{4.00} & \multicolumn{1}{c|}{25.09}                                                                  & \multicolumn{1}{c|}{1317}                                                                    & \multicolumn{1}{c|}{122}        & \multicolumn{1}{c|}{1195}                                                      & \multicolumn{1}{c|}{40}                                                      & \multicolumn{1}{c|}{351}                                                         & \multicolumn{1}{c|}{0.33}                 & \multicolumn{1}{c|}{0.29}                                                               & \multicolumn{1}{c|}{0.12}                                                                                                  & 1.89                                                                                                                             \\ \hline
\multicolumn{1}{|c|}{Expert}  & \multicolumn{1}{c|}{3.60} & \multicolumn{1}{c|}{29.49}                                                                  & \multicolumn{1}{c|}{1720}                                                                    & \multicolumn{1}{c|}{142}        & \multicolumn{1}{c|}{1578}                                                      & \multicolumn{1}{c|}{50}                                                      & \multicolumn{1}{c|}{526}                                                         & \multicolumn{1}{c|}{0.35}                 & \multicolumn{1}{c|}{0.33}                                                               & \multicolumn{1}{c|}{0.10}                                                                                                  & 1.89                                                                                                                             \\ \hline
                              &                           &                                                                                             &                                                                                             &                                &                                                                               &                                                                              &                                                                                  &                                           &                                                                                         &                                                                                                                            &                                                                                                                                  \\ \hline
\multicolumn{1}{|c|}{}        & \multicolumn{2}{c|}{Teacher Words}                                                                                      & \multicolumn{3}{c|}{Teacher Utterances}                                                                                                                                                                      & \multicolumn{2}{c|}{Child Responses}                                                                                                                            & \multicolumn{2}{c|}{\begin{tabular}[c]{@{}c@{}}Proportion of \\ Teacher Utterances \\ Followed by a \\ Child Response\end{tabular}} & \multicolumn{1}{c|}{\begin{tabular}[c]{@{}c@{}} Percentage\end{tabular}} & \multicolumn{1}{c|}{\begin{tabular}[c]{@{}c@{}} Lexical \\Diversity  \end{tabular}}                                                                                                                       \\ \hline
\multicolumn{1}{|c|}{}        & \multicolumn{1}{c|}{MLU}  & \multicolumn{1}{c|}{\begin{tabular}[c]{@{}c@{}}Mean \\ Words \\ per Minute\end{tabular}} & \multicolumn{1}{c|}{\begin{tabular}[c]{@{}c@{}}Total \\ Teacher \\ Utterances\end{tabular}} & \multicolumn{1}{c|}{Questions} & \multicolumn{1}{c|}{\begin{tabular}[c]{@{}c@{}}Non-\\ Questions\end{tabular}} & \multicolumn{1}{c|}{\begin{tabular}[c]{@{}c@{}}To \\ Questions\end{tabular}} & \multicolumn{1}{c|}{\begin{tabular}[c]{@{}c@{}}To Non-\\ Questions\end{tabular}} & \multicolumn{1}{c|}{Questions}            & \multicolumn{1}{c|}{\begin{tabular}[c]{@{}c@{}}Non-\\ Questions\end{tabular}}           & \multicolumn{1}{c|}{Questions}                                                                                            & \begin{tabular}[c]{@{}c@{}}Mean Lexical\\Diversity\\per Minute\end{tabular} \\ \hline
\multicolumn{1}{|c|}{Whisper} & \multicolumn{1}{c|}{4.88} & \multicolumn{1}{c|}{48.13}                                                                  & \multicolumn{1}{c|}{2071}                                                                   & \multicolumn{1}{c|}{481}       & \multicolumn{1}{c|}{1590}                                                      & \multicolumn{1}{c|}{160}                                                      & \multicolumn{1}{c|}{356}                                                         & \multicolumn{1}{c|}{0.33}                 & \multicolumn{1}{c|}{0.22}                                                               & \multicolumn{1}{c|}{0.22}                                                                                                  & 10.47                                                                                                                             \\ \hline
\multicolumn{1}{|c|}{Expert}  & \multicolumn{1}{c|}{4.72} & \multicolumn{1}{c|}{53.67}                                                                  & \multicolumn{1}{c|}{2388}                                                                   & \multicolumn{1}{c|}{565}       & \multicolumn{1}{c|}{1823}                                                      & \multicolumn{1}{c|}{222}                                                     & \multicolumn{1}{c|}{502}                                                         & \multicolumn{1}{c|}{0.39}                 & \multicolumn{1}{c|}{0.28}                                                               & \multicolumn{1}{c|}{0.22}                                                                                                  & 11.23                                                                                                                             \\ \hline
\end{tabular}
\caption{Teacher and Child Language Features and Responses from Whisper and Expert Transcriptions}
\label{table:large}
\end{table*}

\begin{table*}[]
\centering
\begin{tabular}{|l|ll|ll|ll|ll|ll|ll|}
\hline
\multirow{2}{*}{\begin{tabular}[c]{@{}l@{}}Absolute ICCs \\ (WSW vs. Expert)\end{tabular}}         & \multicolumn{2}{l|}{\begin{tabular}[c]{@{}l@{}}Question \\ Utterances\\ per minute\end{tabular}} & \multicolumn{2}{l|}{\begin{tabular}[c]{@{}l@{}}Non-Question \\ Utterances \\ per minute\end{tabular}} & \multicolumn{2}{l|}{\begin{tabular}[c]{@{}l@{}}Response to \\ Question \\ Utterances \\ per minute\end{tabular}} & \multicolumn{2}{l|}{\begin{tabular}[c]{@{}l@{}}Total Response \\ Proportion\end{tabular}} & \multicolumn{2}{l|}{Overall MLU}              & \multicolumn{2}{l|}{\begin{tabular}[c]{@{}l@{}}Lexical \\ Diversity\end{tabular}} \\ \cline{2-13} 
                                                                                                   & \multicolumn{1}{l|}{Teacher}                           & Child                                   & \multicolumn{1}{l|}{Teacher}                                 & Child                                  & \multicolumn{1}{l|}{Teacher}                                      & Child                                        & \multicolumn{1}{l|}{Teacher}                           & Child                            & \multicolumn{1}{l|}{Teacher} & Child          & \multicolumn{1}{l|}{Teacher}                       & Child                        \\ \hline
\textbf{\begin{tabular}[c]{@{}l@{}}WSW2.0 ESC (Egocentric \\ Speaker Classification~\cite{feng2024egocentric})\end{tabular}} & \multicolumn{1}{l|}{0.975}                             & \textbf{0.964}                          & \multicolumn{1}{l|}{\textbf{0.900}}                          & 0.741                                  & \multicolumn{1}{l|}{\textbf{0.642}}                               & \textbf{0.830}                               & \multicolumn{1}{l|}{\textbf{0.932}}                    & 0.856                            & \multicolumn{1}{l|}{0.985}   & 0.713          & \multicolumn{1}{l|}{\textbf{0.854}}                & \textbf{0.863}               \\ \hline
\begin{tabular}[c]{@{}l@{}}WSW1.0 VTC (Voice Type \\ Classifier~\cite{lavechin20_interspeech})\end{tabular}                      & \multicolumn{1}{l|}{0.975}                             & 0.574                                   & \multicolumn{1}{l|}{0.850}                                   & \textbf{0.822}                         & \multicolumn{1}{l|}{0.244}                                        & 0.780                                        & \multicolumn{1}{l|}{0.690}                             & \textbf{0.886}                   & \multicolumn{1}{l|}{0.988}   & \textbf{0.872} & \multicolumn{1}{l|}{0.675}                         & 0.832                        \\ \hline
\end{tabular}
\caption{Intraclass Correlations Comparison between integrated ESC(wav2vec2) and VTC vs. Expert with Whisper large-v2 in WSW. (Utterances were automatically aligned by WSW framework)}
\label{table:3}
\end{table*}

In WSW 2.0, we compute MLU as the number of words per utterance. 
For teacher speech, Whisper yielded a mean MLU of $4.88 (SD=0.92)$ versus $4.72 (SD=0.93)$ for the expert. For children, the Whisper mean MLU was $4.00 (SD=0.58)$, compared to $3.60 (SD= 0.60)$ for the expert (standard deviations were calculated at the clip level). These values suggest high agreement for teachers and moderate agreement for children between Whisper and the human expert in estimating speech complexity as indexed by MLU (see TABLE~\ref{table:large}). Absolute agreement intraclass correlation analyses  provide additional evidence for agreement. The teacher Overall MLU intraclass correlation (ICC) is 0.99, and the child Overall MLU ICC is 0.74.

\subsubsection{Speech Rate} Both Whisper and expert annotations reveal that teachers have an MLU roughly one-forth higher than children. However, teachers also speak at more than twice the rate of children (words per minute) and produce more utterances overall. Whisper suggests a teacher-to-child utterance ratio of $(1.57 [2071/1317])$, slightly higher than the $(1.39 [2388/1720])$ observed by the expert.
These findings suggest WSW 2.0’s potential to quantify a key, validated index of teacher and child language in real-world classroom environments.

\subsubsection{Questions and Non-Questions}
To evaluate the classification of utterances as either questions or non-questions, we compare the presence of question marks in Whisper and expert transcriptions. The data show close alignment (see TABLE~\ref{table:large}). Both Whisper (0.12) and the expert (0.10) indicate that about one-tenth of child utterances are questions, while teachers pose questions in approximately one-fifth of their utterances (0.22 for both Whisper and expert). These proportions suggest robust performance in discriminating question versus non-question utterances under noisy preschool conditions.


\subsubsection{Responses}
Responses are defined as an utterance occurring within 2.5 seconds of the offset of the partner's utterance (see TABLE~\ref{table:large}). Whisper and expert transcriptions converge on similar rates for teacher responses both to child questions and non-questions and child responses to teacher questions, all of which show values around one-third. In contrast, child responses to teacher non-questions range between one-fifth (Whisper) and one-forth (expert).

\subsubsection{Lexical Diversity}
In addition to MLU, we compute mean lexical diversity (LD) to quantify the number of unique words used by teachers and children per minute. Teachers produced  substantially higher LD than children (10.47 vs. 1.89 for Whisper, 11.23 vs. 1.89 for the expert). Notably, Whisper and the expert yield nearly identical LD estimates for children, while teacher LD shows only a modest difference (10.47 vs. 11.23).

\subsubsection{Intraclass Correlations (ICCs)}
In preparing to compute absolute agreement ICCs, we transformed the total number of teacher and child questions (see TABLE~\ref{table:large}) to a rate per minute (see TABLE~\ref{table:3}). We also calculated the total rate per minute of teacher responses to all child utterances (question and non-question utterances) and the total rate per minute of child responses to teacher utterances. 

TABLE~\ref{table:3} reflects automated integration and alignment of speaker classification outcomes (obtained from ESC and VTC) with Whisper (large-v2) transcription results, enabling a direct comparison with expert coding. The first row of TABLE~\ref{table:3} compares the current WSW 2.0 to expert coding. Child ICCs ranged from 0.71 (overall MLU) to 0.96 (question asking rate). Teacher ICCs ranged from 0.64 (response to question rate) to 0.99 (overall MLU). The results underscore WSW 2.0’s capacity to reliably capture key indicators of child and teacher language functioning.


  TABLE V contrasts two WSW approaches with expert coding. While the first row reports WSW 2.0 results, the second row reports WSW1.0 VTC (Voice Type
Classifier) results. ICC values are typically higher for WSW 2.0 (range 0.64 - 0.98) than for WSW 1.0 (range 0.40 - 0.98). Of note, child MLU ICCs were higher in WSW 1.0 (.87) than WSW 2.0 (.74). The results illustrate continuing improvements in the capacity of machine learning pipelines to accurately capture child speech in noisy, everyday environments. 
Higher WSW 2.0 ICC values highlight the enhancements introduced by ESC(wav2vec2), particularly in capturing child language features.

\subsection{Year Scale Data Analysis}

\subsubsection{Dataset}
To illustrate feasibility for longitudinal research, we applied WSW~2.0 with large-v3 to a dataset from one classroom over two academic-years. One class from the 2022–2023 year (denoted 2223; C3) included 159 recordings totaling 685.6 hours across 15 observation dates, while the 2023–2024 year (2324; C5) included 207 recordings totaling 906.8 hours across 20 observation dates. The two classes contain a total of 27 children (14 with hearing loss). Altogether, WSW~2.0 processed 642,564 utterances for 2223 and 843,970 utterances for 2324 in roughly two weeks of computational time. This large-scale deployment underscores the framework’s efficiency and practicality for extended monitoring of preschool language interactions over multiple years.


\section{Discussion}
Preschool classrooms serve as a vital developmental context for children aged 3 to 5 in the United States. Previous research has repeatedly shown that the characteristics of teachers' speech, such as frequent questioning and a longer mean utterance length (MLU), are related to emerging language skills of children \cite{hadley2022meta,walsh2018we}. Likewise, children’s own speech production, including phonemic complexity, can provide a crucial conduit between teacher inputs and child language growth \cite{mitsven2022objectively}. 

Historically, speaker classification and transcription of classroom audio have proven to be a bottleneck for researchers. Expert transcription remains both time-consuming and costly (often requiring multiple hours of expert labor to transcribe just a fraction of recording time). Recent efforts have begun to explore automated strategies for classroom audio analysis \cite{dutta2022activity,seven2024capturing}. Here, WSW~2.0 advances this line of work by integrating a wav2vec2-based speaker classification module and Whisper-based transcription to automate both the “who” (teacher vs. child) and “what” (spoken content) of classroom discourse.

The results for speaker classification suggest moderate-to-high reliability, with teachers classified more accurately than children likely reflecting the distinct acoustic qualities of adult speech. Likewise, word error rates (WER) were lower than might be expected in a noisy preschool context, particularly for teacher-recorded audio. Critically, rather than providing only standard reliability metrics, we also assessed the reliability of key language features. Our findings indicate that automated estimates of teacher and child MLU, speaking rate, question usage, and responsiveness are aligned with expert annotations, highlighting the applicability of WSW~2.0 for substantive classroom research questions.

We present preliminary findings from 35 longitudinal observations (1,592 hours of audio recordings) from two cohorts of an inclusion classroom containing 3-4-year-olds. We found that child MLU increased over observations, and showed substantial variation between children.  Mean child MLU showed associations with assessed receptive and expressive language ability abilities (as measured by the PLS-5). Strikingly, children who had higher mean teacher MLU exhibited larger increases in their own MLU over observations. In sum, the complexity of teacher language heard by individual children was associated with the complexity of the language produced by those children and the rate of increase in the complexity of the child’s speech. Our results suggest how specific child-teacher interactions  are associated with child language development. 



Overall, the findings underscore the promise of automated methods for accelerating research on early childhood language interactions. By minimizing the need for labor-intensive manual transcription, WSW~2.0 enables researchers and practitioners to scale data collection and focus on deeper analyses of how language in the preschool classroom influences children’s developmental trajectories.

\section{Conclusion}
The current study presents WSW~2.0, a framework that combines high-quality worn audio recorders, a wav2vec2-based speaker classifier, and Whisper-based transcription to automate the analysis of teacher-child speech in preschool classrooms. The system yields robust agreement with expert annotations on a variety of substantive metrics, including mean length of utterance, word error rates, and question-response patterns. After validating on a 210-minute dataset, we further demonstrated its scalability by applying the pipeline to over 1,592 hours of recordings, producing hundreds of thousands of transcribed utterances for in-depth examination.

These results illustrate a potential breakthrough in researchers’ ability to capture naturally occurring classroom interactions at scale. Automated speaker classification and transcription not only saves human resources but also open doors to increasingly sophisticated investigations of associations between teacher language and child language development. Moving forward, refinements to improve child speech recognition, handle overlapping talk, and measure more advanced constructs of alignment or semantic complexity are all promising directions~\cite{shivakumar2022end}. Such expansions could deepen our understanding of how subtle features of adult-child interactions facilitate or constrain language learning in the early years. Ultimately, we envision that automated approaches like WSW~2.0 will prove integral to both basic research on language acquisition and the design of interventions that promote equitable educational outcomes for young children.

\section*{Acknowledgement}
This work was supported by the National Institute on Deafness and Communication Disorders (R01DC018542), the National Science Foundation (2150830), and the Simons Foundation Autism Research Initiative (SFI-AR-HUMAN-0000411501).

\end{document}